\documentclass[11pt,twoside]{article}
\usepackage{asp2010}

\resetcounters

\begin{document}

\title{Constraining Variable High Velocity Winds from Broad Absorption Line Quasars with Multi-Epoch Spectroscopy}
\author{Daryl Haggard,$^1$ Kenza S. Arraki,$^2$ Paul J. Green,$^3$ Tom Aldcroft,$^3$ and Scott F. Anderson$^4$
\affil{$^1$Center for Interdisciplinary Exploration and Research in Astrophysics (CIERA), 
Northwestern University, Evanston, IL 60208, USA} 
\affil{$^2$Department of Astronomy, New Mexico State University, Las Cruces, NM 88003, USA}
\affil{$^3$Harvard-Smithsonian Center for Astrophysics, Cambridge, MA 02138, USA} 
\affil{$^4$Department of Astronomy, University of Washington, Seattle, WA 98195, USA}} 

\begin{abstract}
Broad absorption line (BAL) quasars probe the high velocity gas ejected by luminous accreting black holes. BAL variability timescales place constraints on the size, location, and dynamics of the emitting and absorbing gas near the supermassive black hole. We present multi-epoch spectroscopy of seventeen BAL QSOs from the Sloan Digital Sky Survey (SDSS) using the Fred Lawrence Whipple Observatory's 1.5m telescope's FAST Spectrograph. These objects were identified as BALs in SDSS, observed with {\em Chandra}, and then monitored with FAST at observed-frame cadences of 1, 3, 9, 27, and 81 days, as well as 1 and 2 years. We also monitor a set of non-BAL quasars with matched redshift and luminosity as controls. We identify significant variability in the BALs, particularly at the 1 and 2 year cadences, and use its magnitude and frequency to constrain the outflows impacting the broad absorption line region. 
\end{abstract}

\section{Introduction}

Studies of quasar mass outflows are rapidly reshaping our understanding of physics near the SMBH. Though the $\sim$parsec scale emitting/absorbing regions near SMBHs will remain spatially unresolved for the foreseeable future (milliarcsec at $z$$\sim$0.01), time-dependent calculations \citep{Proga00} show that disk wind instabilities result in a filamentary substructure, which gives rise to column density variations along our sightline. Combined with rotation, spectroscopic variability timescale information {\em can} therefore constrain absorber size scales, {\em viz.} $a_V\sim v_{trans}\Delta t$. Though outflows likely accompany all luminous accretion disks, broad absorption line (BAL) quasars provide the most dramatic examples (e.g., Hall et al. 2011) --- their massive outflows display P-Cygni profiles that span velocities to $\sim0.3 c$ and are visible in the spectra of $15-40$\% of optically-selected quasars \citep[e.g.,][]{Dai08,Gibson09,Allen11}.

We present a BAL spectroscopic variability campaign \citep{Haggard11a} that includes seventeen $i < 17$ BAL QSOs from the SDSS with {\em Chandra} X-ray flux and spectral information, as well as {${f}_x$/${f}_{opt}$}\, (another key indicator of absorption). We draw from the huge population of non-BAL SDSS QSOs to derive a sample of 10 control QSOs, well-matched in $i$ mag and redshift (and thus in absolute mag $M_i$). Monitoring is performed with the FAST spectrograph at the Fred Lawrence Whipple Observatory 1.5m Tillinghast telescope. Since the variability timescale is not known, we sample geometrically across 3 nights, then one night each at 9, 27, and 81 days, and 1, 2, and 6 years (the 6 year cadence observations are scheduled for early 2012). 

\section{Results \& Discussion}

\begin{figure}
\includegraphics[width=0.54\textwidth]{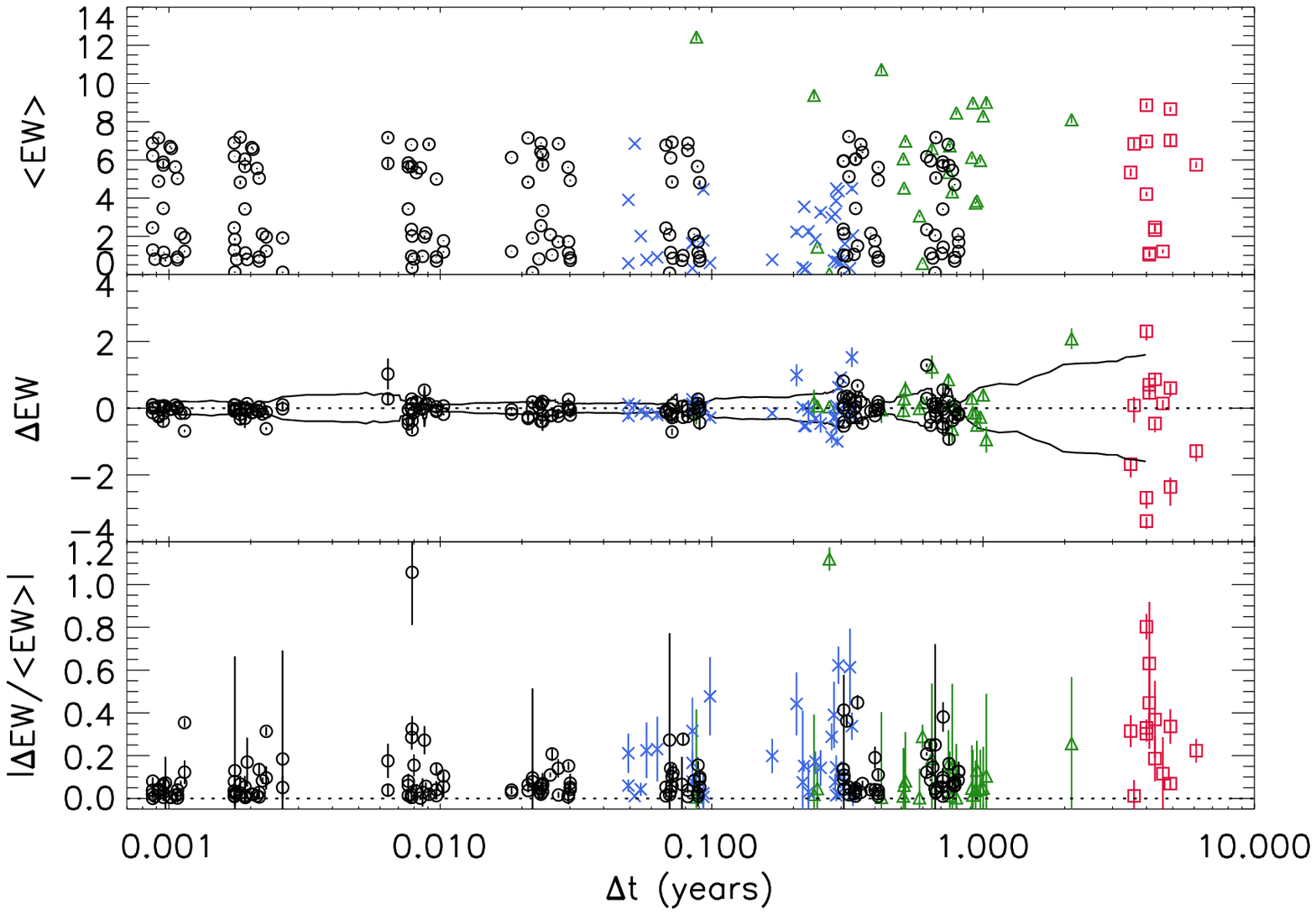}
\includegraphics[width=0.52\textwidth]{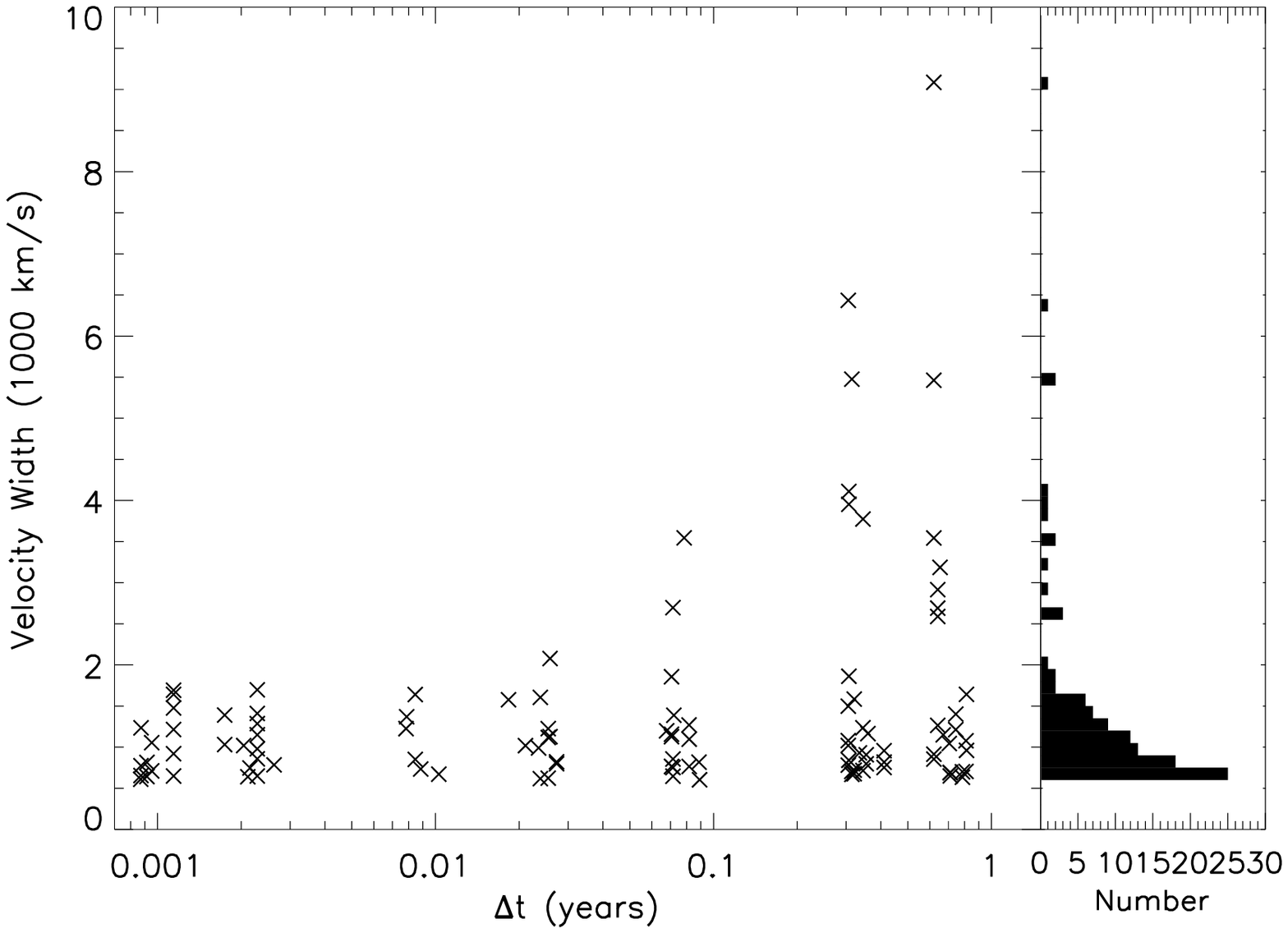}
\vspace{-0.5cm}
\caption{{\footnotesize 
({\it Left}\,) Average CIV BAL equivalent width ($<$EW$>$, units: 1000 km/s; top), the change in EW ($\Delta$EW, units: 1000 km/s; middle), and the magnitude of the fractional change in EW (bottom) against the rest-frame time between epochs ($\Delta \mathrm{t}$). Data are from our FAST study \citep[circles,][]{Haggard11a}, and several previous studies (triangles, Barlow et al. 1993; crosses, Lundgren et al. 2007; squares, Gibson et al. 2008). Solid lines (middle) mark the square root of the unbiased sample variance from a sliding window of 15 time-ordered entries. The envelope of $\Delta \mathrm{EW}$\, expands with time. The 6 year cadence ($\Delta \mathrm{t} = 2$ years rest-frame at z $=$ 2) scheduled for early 2012 will fill the crucial gap at 1 $\le \Delta \mathrm{t} \le 3$ years.
({\it Right}\,) The total width of each CIV variable region displayed versus rest-frame $\Delta \mathrm{t}$. The sub-panel displays a histogram of the number of points with a given velocity width. 
}}
\label{bal_trends}
\vspace{-0.3cm}
\end{figure}

Our ensemble sample is represented in Figure \ref{bal_trends}, where we show variations in equivalent width and velocity width. We observe variability in more than one BAL (CIV and SiIV) and most strongly at multi-year times scales (Fig. \ref{bal_trends}). In early 2012, we will complete the 6-year cadence, filling a crucial gap, and include spectroscopy from SDSS to extend our baseline to {\lower0.8ex\hbox{$\buildrel >\over\sim$}} 10 years ({\lower0.8ex\hbox{$\buildrel >\over\sim$}} 5 years rest-frame). We are also developing an analysis pipeline with an eye toward future spectroscopic variability studies, in anticipation of the explosion in time-domain imaging (e.g., PanSTARRS and LSST).

\end{document}